# Movie Editing and Cognitive Event Segmentation in Virtual Reality Video

ANA SERRANO, Universidad de Zaragoza, I3A
VINCENT SITZMANN, Stanford University
JAIME RUIZ-BORAU, Universidad de Zaragoza, I3A
GORDON WETZSTEIN, Stanford University
DIEGO GUTIERREZ, Universidad de Zaragoza, I3A
BELEN MASIA, Universidad de Zaragoza, I3A

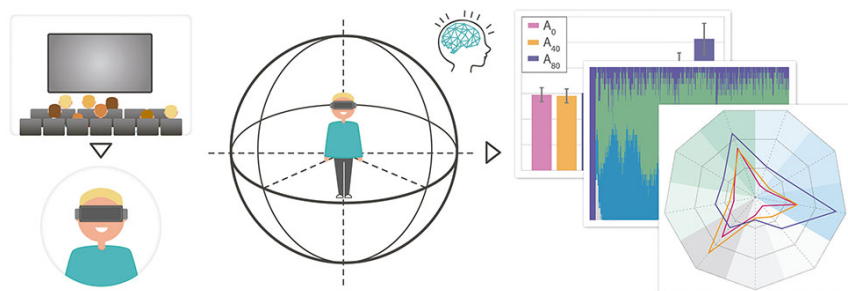

Fig. 1. Different from traditional cinematography, watching a VR movie offers viewers control over the camera. This poses many questions as to what editing techniques can be applied in this new scenario. We investigate the perception of continuity while watching edited VR content, gathering eye tracking data from many observers. We rely on recent cognitive studies, as well as well-established cinematographic techniques, to provide an in-depth analysis of such data, and to understand how different conditions affect viewers' gaze behavior.

Traditional cinematography has relied for over a century on a well-established set of editing rules, called continuity editing, to create a sense of situational continuity. Despite massive changes in visual content across cuts, viewers in general experience no trouble perceiving the discontinuous flow of information as a coherent set of events. However, Virtual Reality (VR) movies are intrinsically different from traditional movies in that the viewer controls the camera orientation at all times. As a consequence, common editing techniques that rely on camera orientations, zooms, etc., cannot be used. In this paper we investigate key relevant questions to understand how well traditional movie editing carries over to VR, such as: Does the perception of continuity hold across edit boundaries? Under which conditions? Does viewers' observational behavior change after the cuts? To do so, we rely on recent cognition studies and the event segmentation theory, which states that our brains segment continuous actions into a series of discrete, meaningful events. We first replicate one of these studies to assess whether the predictions of such theory can be applied to VR. We next gather gaze data from viewers watching VR videos containing different edits with varying parameters, and provide the first systematic analysis of viewers' behavior and the perception of continuity in VR. From this analysis we make a series of relevant findings; for instance, our data suggests that predictions from the cognitive event segmentation theory are useful guides for VR editing; that different types of edits are equally well understood in terms of continuity; and that spatial misalignments between regions of interest at the edit boundaries favor a more exploratory behavior even *after* viewers have fixated on a new region of interest. In addition, we propose a number of metrics to describe viewers' attentional behavior in VR. We believe the insights derived from our work can be useful as guidelines for VR content creation.

## 1 INTRODUCTION

Movies are made up of many different camera shots, usually taken at very different times and locations, separated by cuts. Given that the resulting flow of information is usually discontinuous in space, time, and action, while the real world is not, it is somewhat surprising that the result is perceived as a coherent sequence of events. The key to maintaining this illusion lies in how these shots are edited together, for which filmmakers rely on a system called *continuity editing* [Bordwell et al. 1997; O'Steen 2009]. Although other techniques exist to link shots together, such as the fade-out, fade-in, or dissolve, approximately 95% of editing boundaries are cuts [Cutting 2004], which directly splice two camera frames.

The goal of continuity editing in movies is then to create a sense of situational continuity (a sequence of shots perceived as a single event), or discontinuity (transitions from one event or episode to another). Professional editors have developed both a strong sense of rhythm and a solid intuition for editing together camera shots, making the cuts "invisible". For instance, spatial continuity is maintained largely by the 180° rule, stating that the camera should not cross the axis of action connecting two characters in a shot, while continuity of action is achieved by starting the action in one shot and immediately continuing it in the shot after the cut. Some authors have proposed partial theories to explain why these edited shots are perceived as a continuous event. For example, the 180° rule creates a virtual stage where the action unfolds [Bordwell et al. 1997], while mechanisms to process and track biological motion may mask an action cut [Smith and Henderson 2008].

However, the higher level cognitive processes that make continuity editing work are not yet completely understood. What is understood, though, is that a core component of spatial perception is our ability to segment a whole into parts [Biederman 1987]. Recent cognitive and neuroscience research indicates that a similar segmentation also occurs in the temporal domain, breaking up a continuous activity into a series of meaningful events. This has lead to the development of the *event segmentation* theory [Kurby and Zacks 2008; Reynolds et al. 2007; Zacks and Swallow 2007], which postulates that our brains use this discrete representation to predict the immediate course of events, and to create an internal, interconnected representation in memory. New events are registered whenever a change in action, space, or time, occurs. Based on this theory, recent works have explored how continuity is perceived in movies, across different types of cuts [Cutting 2014; Magliano and Zacks 2011; Zacks et al. 2010]. Interestingly, it seems that the predictive process suggested by event segmentation theory is consistent with common practice by professional movie editors.

In this work, we investigate continuity editing for *virtual reality videos*[1]. Virtual reality (VR) content is intrinsically different from traditional movies in that viewers now have partial control of the camera; while the position of the viewer within the scene is decided during acquisition, the orientation is not. This newly-gained freedom of users, however, renders many usual techniques, such as camera angles and zooms, ineffective when editing the movie. Nevertheless, new degrees of freedom for content creators are enabled, and fundamental questions as to what aspects of the well-established cinematographic language apply to VR should be revisited (Fig. 1). In particular, we seek to investigate answers to the following key questions:

- Does continuity editing work in VR, i.e., is the perception of events in an edited VR movie similar to traditional cinematography?
- A common, safe belief when editing a VR movie is that the regions of interest should be aligned before and after the cut. Is this the only possible option? What are the consequences of introducing a misalignment across the cut boundaries?
- Are certain types of discontinuities (cuts) in VR favored over others? Do they affect viewer behavior differently?

We use a head mounted display (HMD), equipped with an eye tracker, and gather behavioral data of users viewing different VR videos containing different types of edits, and with varying parameters. We first perform a study to analyze whether the connections between traditional cinematography and cognitive event segmentation apply to immersive VR (Sec. 4). To this end, we replicate a recent cognitive experiment, previously carried out using traditional cinematographic footage [Magliano and Zacks 2011], using instead a VR movie. Our results show similar trends in the data, suggesting that the same key cognitive mechanisms come into play, with an overall perception of continuity across edit boundaries.

We further analyze continuity editing for VR, exploring a large parameter space that includes the type of edit from the cognitive point of view of event segmentation, the number and position of regions of interest before and after the cut, or their relative alignment across the cut boundary (Sec. 5). We propose and leverage a series of novel metrics that allow to describe viewers' attentional behavior in VR (Sec. 6), including fixations on a region of interest, alteration of gaze after a cut, exploratory nature of the viewing experience, and a state sequence analysis of the temporal domain according to whether the viewer is fixating on a region or performing saccadic movements.

Our analyses reveal some findings that can be relevant for VR content creators and editors: for instance, that predictions from the cognitive event segmentation theory seem to be useful guides for VR editing; that different types of edits are equally well understood in terms of continuity; how the dependence of the time to convergence to a region of interest after a cut is not linear with the misalignment between regions of interest at the cut, but rather appears to follow an exponential trend; or how spatial misalignments between regions of interest at the edit boundaries elicit a more exploratory behavior even *after* viewers have fixated on a new region of interest. We believe our work is the first to empirically test the connections between continuity editing, cognition, and narrative VR, as well as to look into the problem of editing in VR, in a systematic manner. In addition, we provide all our eye tracking data, videos, and code to help other researchers build upon our work.[2]

## 2 RELATED WORK

*Tools for editing.* Creating a sequence of shots from raw footage while maintaining visual continuity is hard, especially for novice users [Davis et al. 2013]. Automatic cinematography for 3D environments was proposed by He at al. [1996], encoding film *idioms* as hierarchical finite state machines, while Christianson et al. [1996] proposed a declarative camera control system. Many other different tools have been devised to help in the editing process, usually leveraging the particular characteristics of specific domains such as 3D animations [Galvane et al. 2015], interview videos [Berthouzoz et al. 2012], narrated videos [Truong et al. 2016], classroom lectures [Heck et al. 2007], group meetings [Ranjan et al. 2008], egocentric footage [Lu and Grauman 2013], or multiple social cameras [Arev et al. 2014]. Jain et al. [2014] proposed a gaze-driven, re-editing system for retargeting video to different displays. More recently, Wu and Christie [2015] created a language to define camera framing and shot sequencing. Other methods focusing on camera placement and planning can be found in the work of Christie et al. [Christie et al. 2005]. All these tools have been designed for traditional, two-dimensional viewing experiences, where the spectator sits passively in front of a screen. In contrast, our goal is to analyze continuity editing for virtual reality videos.

---

[1]In this work we deal with 360° movies; throughout the text we will use the terms VR and 360° movies interchangeably.

[2]http://webdiis.unizar.es/~aserrano/projects/VR-cinematography

*Continuity and cognition.* Several works have analyzed the effects of edits or cuts from a computer vision perspective (e.g., [Carroll and Bever 1976; Hochberg and Brooks 2006; Smith and Henderson 2008]). Closer to our approach, a few works have analyzed the perception of continuity from a cognitive science point of view. For instance, Cohn's analyses of comic strips [2013] suggest that viewers can build links between frames while maintaining a global sense of the narrative; however, rearranging elements can quickly lead to confusion. Some researchers argue that our perception of reality is a very flexible process, and this flexibility allows us to adapt and perceive edited film as a continuous story [Anderson 1996; Cutting 2004]. Smith [2012] performed an empirical study to understand how continuity editing aligns with our perceptual abilities, identifying the role of visual attention in the perception of continuity between edits. In our work, we explore the recent theory of *event segmentation* [Kurby and Zacks 2008; Reynolds et al. 2007; Zacks 2010; Zacks and Swallow 2007], and analyze its connections with continuity editing for VR.

## 3 BACKGROUND ON EVENT SEGMENTATION

We present here a brief summary of the event segmentation theory, and refer the reader to the original publications for a more thorough explanation [Kurby and Zacks 2008; Reynolds et al. 2007; Zacks 2010; Zacks and Swallow 2007]. Recent research suggests that event segmentation is an *automatic* key component of our perceptual processing, reducing a continuous flow of activity into a hierarchical, discrete set of events. The advantages of this strategy are twofold: First, it is very efficient in terms of internal representation and memory. Second, it provides a much easier way to think about events in relation to one another. It can be seen as the time equivalent to the well-known spatial segmentation in vision, where we segment an object (e.g., a car) into many components such as wheels, chassis, engine, etc.

This discrete mental representation is used as a basis for predicting the immediate course of events: a person walking down the street will continue to do so, or somebody will answer a question when asked. When these predictions are violated, it is an indication of a new (discrete) event; in other words, it seems that unexpected changes lead to the perception of an event boundary. More precisely, the event segmentation theory assumes that new events are registered when *changes in action, space, or time*, occur; when this happens, the mechanisms of event segmentation update the mental representation of the event, storing the old one in long-term memory.

This event segmentation theory has recently been tested in the context of film understanding. Some experiments have even recorded brain activity with functional magnetic resonance imaging (fMRI) while watching a movie, and showed that many regions in the cortex underwent substantial changes in response to the situational discontinuities (unexpected changes) introduced by some movie cuts [Magliano and Zacks 2011; Zacks et al. 2010]. An interesting observation follows: *the predictive process suggested by event segmentation theory is consistent with common practice by professional movie editors*, who place cuts to support or break the expectations of event continuity by the viewers [Bordwell et al. 1997]. When a cut introduces a major change, the brain does not try to explain the perceived discontinuity; instead, it adapts to the change, creates a new mental representation, and begins populating it with details [Magliano and Zacks 2011]. This automatic mechanism might be a key process to explain why continuity editing works. The next section explores whether this connection between event segmentation and continuity editing studied in traditional cinema carries over to VR movies, a key question before we can dive into a more detailed investigation. Note that, in the following, we use the term *edit* to refer to a discontinuity between two shots, while *cut* refers to the actual cinematographic implementation (match-on-action, jump cut, etc.) of the edit.

## 4 DOES CONTINUITY EDITING WORK IN VR?

As we have seen in Sec. 3, there is considerable evidence that continuity editing performed in traditional movies may be related to how our brains process events and situational changes, and that this may be the cause why continuity editing has been so successful in conveying the narrative. Therefore, before we analyze specific aspects related to editing in VR movies, we first want to assert that continuity editing applies to VR scenarios. For this purpose, we check whether the connections between event segmentation and edits, which have been identified and analyzed in traditional movies [Magliano and Zacks 2011] also hold in VR movies, where the viewing conditions and the perception of immersion change significantly. This is the goal of the experiment described in this section. We aim to replicate the methodology of recent cognition studies, sharing a similar goal in the contexts of event segmentation [Zacks and Swallow 1976], and film understanding [Magliano and Zacks 2011; Zacks et al. 2010]. We introduce such works and our own experiment in the following paragraphs.

*Types of edits.* Following common practice in film editing, Magliano and Zacks [2011] define a continuity domain along the dimensions of space, time, and action. They then classify edits into three different classes, which we call here $E_1$, $E_2$, and $E_3$:

- $E_1$: edits that are discontinuous in space or time, and discontinuous in action (action discontinuities);
- $E_2$: edits that are discontinuous in space or time, but continuous in action (spatial/temporal discontinuities);
- $E_3$: edits that are continuous in space, time, and action (continuity edits).

We adopt the same taxonomy for edits in this experiment, and in the rest of the paper.

*Cognition studies with traditional movie content.* In these studies [Magliano and Zacks 2011; Zacks et al. 2010], participants watched *The Red Balloon* (a 33-minute, 1956 movie by A. Lamorisse), and were asked to segment the movie into meaningful events by pressing a button. They were asked to do this twice, once defining the *"largest units of natural and meaningful activity"* (coarse segmentation), and once defining the smallest units (fine segmentation); the order of this division was randomized between participants, who first practiced the task on a different, 2.5-minute movie. *The Red Balloon* was presented in 7-to-10-minute sections, to avoid fatigue. Previous to this task, the authors additionally identified all the locations where edits occurred in the movie, and coded each one according to the above categorization: $E_1$, $E_2$, or $E_3$. Based on the principles of film editing discussed in Sec. 3, action discontinuities $E_1$ should have the largest influence on perceived discontinuities, whereas continuity edits $E_3$ should mostly maintain the perceived continuity. The analysis of the data discretized in five-second bins, along with fMRI information, confirmed this predicted trend.

*Replication of the study with VR content.* We followed the same methodology in our VR study. Specifically, we asked seven participants (ages between 21 and 31, three female) to watch *four* publicly available VR movies (see Appendix for details). Their initial task was to mark perceived event boundaries both at coarse and fine scales, similar to the original experiments. Previous to the experiment, the participants watched a training movie[3]. We did not find strong correlations between the fine event segmentation and the edits in the first tested VR movie. This is expected since, different from traditional movies, the time scale of such fine perceived events is about one order of magnitude smaller than the average VR shot (seconds vs. tens of seconds). Therefore in the other three movies we only asked them to mark events at coarse scale; we analyze this data only in the rest of the paper.

Participants watched the movies while seated, wearing an Oculus Rift. We chose these particular movies among several candidates since: i) like *The Red Balloon*, they are narrative movies with a rich enough structure; ii) they last less than eight minutes, which falls within the range fixed by previous studies to avoid fatigue; iii) their average shot lasts about 20 seconds, close to the average we obtained from analyzing several 360° movies; and iv) they contain edits of all three kinds. Fig. 2 (top) shows representative frames of all three types of edits for one of these movies: *Star Wars - Hunting of the Fallen.*

*Insights.* To investigate the relation between the edits and the perceived event segmentation, we identified the location and type of each edit, and binned all the event boundaries marked by the participants within a ±3 second window, centered at the edit. As expected, some perceived boundaries were not linked to edits, but to new events or actions within a shot (e.g., a new actor entering a scene, or the start of a conversation). Fig. 2 (bottom) shows the results for all four movies, grouped by edit category. Our findings show similarities with the previous studies on traditional cinematography [Magliano and Zacks 2011; Zacks et al. 2010]. First, action discontinuities dominate event segmentation, and are therefore the strongest predictors of event boundaries. Second, continuity edits succeed in maintaining a sense of continuity of action, even across the edit boundaries. It thus appears that the key cognitive aspects of traditional movie editing that make it work so well carry over to 360° immersive narrative movies. To our knowledge, this is the first time that this has been empirically tested.

[3]https://youtu.be/fz88kpRNTqM

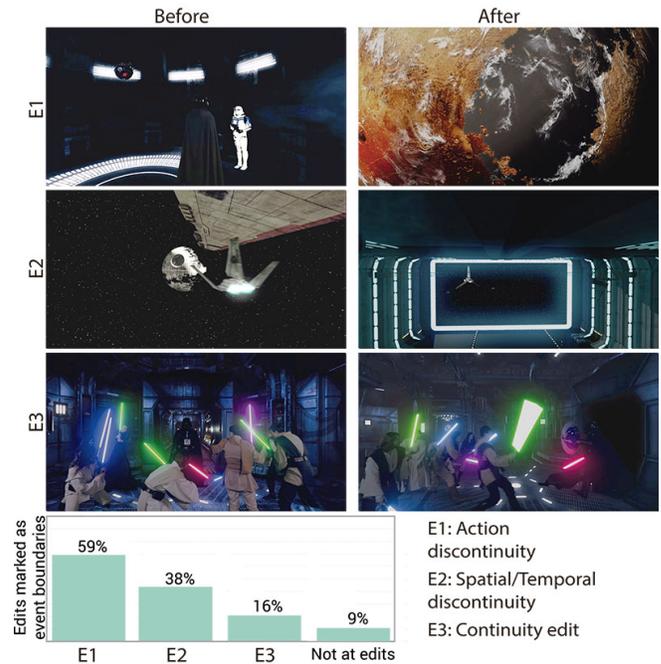

Fig. 2. **Top:** Representative frames of the 360° movie *Star Wars - Hunting of the Fallen*, before (left) and after (right) an edit for each of the three types of edits (original video property of CubeFX (http://cubefx.cz); video and images used with permission). Top row: action discontinuity ($E_1$). The frame before the edit is not related with the frame after the edit; there is a complete change of action, space, and time. Middle row: spatial discontinuity ($E_2$). The edit follows the action by the spaceship (match-on-action), but changes the location; the action is therefore the element connecting the frames before and after the edit. Bottom row: continuity edit ($E_3$). The same scene is depicted before and after the edit with a continuity in space, time, and action; only the camera angle changes. **Bottom:** Results of our coarse segmentation test, showing the percentage of edits of each type marked as an event boundary by subjects, and normalized by the number of occurrences of each type of edit. $E_1$ action discontinuities dominate event segmentation, while $E_3$ continuity edits maintain the perceived continuity of the event. There is also a small percentage of event boundaries that were marked not at edits. These findings match results of similar studies in traditional cinematography, and suggest that movie editing rules and common practice can be in general applied to narrative VR as well.

## 5 MEASURING CONTINUITY IN VR

After confirming in the previous section that the perception of continuity is maintained across edit boundaries in VR narrative content, we now perform a second, in-depth study to assess *how the different parameters that define an edit in VR affect the viewers' behavior after the edit takes place*. Given the high dimensionality of this space, we focus on four main parameters (or *variables of influence*), which are: the type of edit, for which we follow the cognitive taxonomy described in previous sections; the degree of misalignment of the regions of interest (ROIs) before and after the edit; and the number and location of such ROIs, both before and after the edit boundaries. In the following we describe our stimuli, variables of influence, and procedure. Additional details can be found in the Appendix, while a complete table with all the possible combinations of conditions tested can be found in the supplemental material.

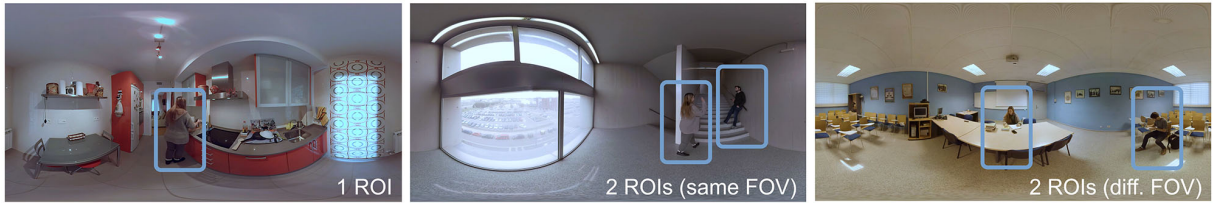

Fig. 3. Representative frames of three of the scenes depicted in our clips: *Kitchen*, *Stairs*, and *Study* (refer to the project page for full videos). From left to right, examples corresponding to the following region of interest (ROI) configurations: 1 ROI, 2 ROIs in the same FOV, and 2 ROIs in different FOV. For clarity, ROIs are marked by a blue box.

### 5.1 Stimuli

Our stimuli are created from 360° monocular videos, professionally captured and stitched by a local company. We choose to use monocular (and not stereo) footage since it is more common among existing capture devices and public repositories (e.g., YouTube 360). The videos depict four different scenarios (*Stairs, Kitchen, Living Room, Study*), with four different actions in each one, totalling sixteen videos ranging from 13 seconds to 2 minutes in length. Fig. 3 shows some representative frames in equirectangular projection. They were captured using two different rigs: a GoPro Omni (a 360-video rig consisting of six GoPro Hero4 cameras), and a Freedom360 3× rig (with three GoPro Hero4 cameras with modified Entaniya 220 lenses). Sound was recorded using a Zoom F8 recorder with wireless microphones.

From these videos, we created a total of 216 *clips*, sampling our parameter space as explained in the following subsection. Each clip is made up of two shots, separated by an edit. Shots are taken from short sequences both within and across the four different scenarios, to maximize variety. Each shot lasts six seconds, to provide enough time to the viewers to understand the actions being shown.

### 5.2 Variables of influence and parameter space

*Type of edit.* We rely on the event segmentation theory, and initially consider the three different types of edits $\{E_1, E_2, E_3\}$, defined along the dimensions of space, time, and action, as introduced in Sec. 3. However, after analyzing seventeen VR movies we have observed that $E_3$ (which essentially refers to a change of viewpoint within the same scene) is rarely used in narrative VR. In these movies, 73% of the edits corresponded to $E_1$, 25% to $E_2$, and only 2% to $E_3$. This differs from traditional movies, where most of the edits are continuity edits ($E_3$) [Magliano and Zacks 2011], and reflects an interesting contrast between the established storytelling techniques for the two media. Due to the rare appearance of $E_3$ edits in VR movies, we remove it from our conditions, and focus on the two most prominent types of edits: $\mathbf{E} = \{E_1, E_2\}$.

For the actual implementation of these edits, we revise traditional cinematography techniques and analyze existing VR videos, and select the most common cuts for each type of edit: For type $E_1$ (discontinuous in action, and in time or space) we use jump cuts, while for $E_2$ (continuous in action, discontinuous in time *or* space) we use compressed time cuts, and match-on-action cuts (see, e.g., [Chandler 2004; Dmytryk 1984]; please refer to the Appendix for a brief explanation of each one). To keep a balanced number of clips for each type of edit, we include twice as many jump cuts (type $E_1$) as match-on-action and compressed time cuts (type $E_2$).

*Alignment of ROIs.* We define the regions of interest (ROIs) as the areas in the 360° frame in which the action takes place[4]. Since the point of view of the camera cannot be controlled by the filmmaker in VR, a common practice among content creators is to simply align ROIs before and after an edit, to make sure that the viewer does not miss important information. However, the exploration of controlled (mis)alignments is interesting for the following reasons: First, the director may want to introduce some misalignment between ROIs for artistic or narrative purposes (e.g., to create tension). Second, the viewer may not be looking at the predicted ROI before the cut, thus rendering the alignment after the edit useless. Third, there might be multiple ROIs within a scene. We therefore test three different alignment conditions: (i) perfect alignment before and after the edit (i.e., 0° between ROIs); (ii) a misalignment that is just within the field of view (FOV) of the HMD[5]; we chose 40° since it is close to the average misalignment in 360° videos found in public repositories; and (iii) a misalignment that is outside the FOV; we chose 80°, since we found that larger values are very rare. We name these conditions $\mathbf{A} = \{A_0, A_{40}, A_{80}\}$.

*ROI configuration.* The control of the viewer over the camera also makes the disposition and number of ROIs in the scene play a key role in gaze behavior. To analyze the ROI configuration before and after the edit, we introduce two variables, $\mathbf{R_b}$ and $\mathbf{R_a}$ respectively. The space of possible configurations is infinite, so to keep the task tractable we test three possibilities for each one: a single ROI ($R_{\{b|a\},0}$), two ROIs both falling within a single FOV ($R_{\{b|a\},1}$), and two ROIs not within the same FOV, i.e., more than 95° apart ($R_{\{b|a\},2}$). Examples of the three configurations are shown in Fig. 3. The possible combinations of $\mathbf{R_b}$ and $\mathbf{R_a}$ yield a total of nine conditions.

---

[4]We manually label ROIs as continuous regions at several keyframes, creating the rest through interpolation. We define the center of each ROI as the centroid of its pixels.
[5]Our Oculus DK2 HMD has a horizontal FOV of 95°, so a 40° misalignment falls just in the periphery of the FOV.

*Summary.* This sampling leads to 2 (types of edit) × 3 (alignments) × 9 (ROI configurations) = 54 different conditions. For each one, we include four different clips, to minimize the effect of the particular scene shown, yielding our final number of 216 stimuli.

### 5.3 Hardware and procedure

We used an Oculus DK2 HMD equipped with a binocular eye tracker from pupil-labs[6], which records data at 120 Hz with a spatial accuracy of 1 degree. We also used a pair of headphones to reproduce stereo sound. Subjects stood up while viewing the video. A total of 49 subjects (34 male, 15 female, $\mu_{age}$ = 25.4 years, $\sigma_{age}$ = 7.7 years) participated in the experiment. All of them reported normal or corrected-to-normal vision. Each subject first carried out the eye tracker calibration procedure. Then, they were shown 36 stimuli from the total of 216, in random order. This randomization was such that no subject viewed two alignment conditions of the same clip, while guaranteeing that each clip was viewed by at least five people. It also avoids potential learning and fatigue effects affecting the results. Following Sitzmann et al. [2016], in order to ensure that the starting condition was the same for all subjects, a gray environment with a small red box was displayed between clips; users had to find it and align their head direction with it, which would launch a new clip after 500 ms. The Unity game engine was used to show the videos, and to record head orientation on the same computer, while eye tracking data was recorded on a second computer. After viewing the clips the experimenter did a debriefing session with the subject. The total time per experiment was around 15 minutes. From the raw gathered data, we performed outlier rejection and then computed *scanpaths*, defined as a temporal sequence containing one gaze sample per frame. More details on these aspects can be found in the Appendix (gaze data processing and outlier rejection), and in the supplemental material (data collection and debriefing). From this data we define, compute, and analyze a series of metrics, as described in the next section.

## 6 HOW DO EDITS IN VR AFFECT GAZE BEHAVIOR?

To obtain meaningful data about viewers' gaze behavior across event boundaries in VR, we first gather additional *baseline data* to compare against. We make the assumption that the higher the gaze similarity between the edited clips and the corresponding (unedited) baseline videos, the higher the perception of continuity; this assumption is similar to previous works analyzing gaze to assess the impact of retargeting and editing operations in images and video [Castillo et al. 2011; Jain et al. 2014]. In the following, we first describe how this baseline data is obtained, then introduce our continuity metrics, and describe the results of our analysis.

### 6.1 Baseline data

In order to capture baseline eye tracking data from the unedited videos, we gathered ten new subjects (nine male, one female, $\mu_{age}$ = 28.1 years, $\sigma_{age}$= 5.2 years) and collected head orientation and gaze data following the procedure described in Sec. 5.3. Videos were watched in random order. We compute the baseline scanpaths for each video from the obtained gaze data as the mean scanpath across users. We show in Fig. 4 the mean scanpath corresponding to one of our videos: We display the temporal evolution of the longitudinal gaze position (0°- 360°), and it shows how viewers' attention is driven towards the ROI moving across the scene.

---
[6]https://pupil-labs.com/

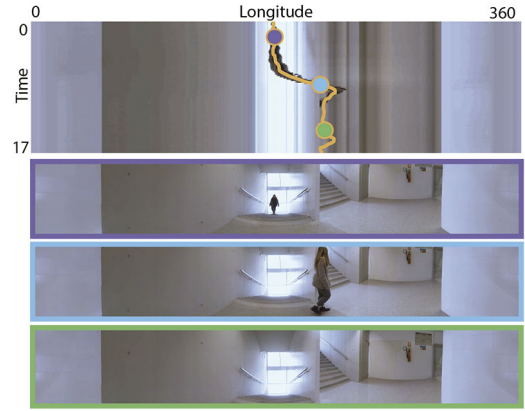

Fig. 4. Subjects mean scanpath for one example video. We plot one scanline per frame of the video, the x-axis showing longitudinal position (0°-360°), and the y-axis time. Superimposed (orange line) we plot the scanpath, showing the temporal evolution of the longitudinal position of the gaze. We also plot the full frame (equirectangular projection) at three key instants that correspond to the three marked temporal instants. Viewers' gaze is clearly directed by the movement of the ROI along time.

To ensure that this data can be used as baseline for our subsequent analyses, we need to ascertain the congruency between subjects. To do so, we rely on a *receiver operating characteristic* curve (ROC) metric, which provides a measure of the *Inter Observer Congruency* (IOC) [Le Meur et al. 2011] over time. First, we aggregate all the users' fixations (please refer to the Appendix for a description of how fixations are computed from gaze data) in two-second windows, and convolve them with a 2D Gaussian of $\sigma$ = 1 degree of visual angle [Le Meur and Baccino 2013], yielding a saliency map for each time window. The corresponding ROC curve is then obtained using a one-against-all approach by leaving out the $i_{th}$ subject: we compute, for each saliency map, the $k\%$ most salient regions, and then calculate the percentage of fixations of the $i_{th}$ subject that fall within those regions. This process is performed for a set of thresholds $k$ = 0%..100%, and the resulting points define each curve. Additionally, we compute the *Area Under the Curve* (AUC) for each window, which provides an easier interpretation of the evolution of the IOC along time (Fig. 5, right). The AUC takes values between 0 (incongruity between users) and 100 (complete congruency). As displayed in Fig. 5, the congruency between subjects remains very high along time. On the left of the figure, the IOC rapidly reaches a value of 1 with $k$ = 2% most salient regions, and remains constant for increasing values of $k$. On the right, the same interpretation from an AUC perspective: all the viewer's fixations fall on average within the 2% regions considered most salient by the rest of the viewers, yielding a very high AUC. This indicates that all the viewers consistently considered the same regions salient. Please, refer to the supplemental material for the results for all our videos.

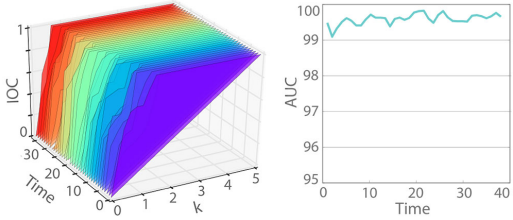

Fig. 5. **Left**: Inter Observer Congruency (IOC) for one of our videos. We compute a ROC curve for each second of the video. **Right**: Temporal evolution of the Area Under the Curve (AUC) calculated for each of the ROC curves. The high values of the IOC and AUC indices indicate that all the viewers consistently considered the same regions salient (refer to the main text for details).

### 6.2 Metrics

Measuring the perceived continuity across edit boundaries in an objective manner is not a simple task, since no predefined metrics exist. We describe here our four different metrics used to analyze gaze behavior after an edit. In addition, to further look for underlying patterns in the users' behavior that our metrics may not capture, we also introduce a *state sequence* analysis.

*Frames to reach a ROI (*framesToROI*).* This is the simplest of our metrics, simply indicating the number of frames after the occurrence of the edit before the observer fixated on a ROI. It is indicative of the time taken to converge again to the main action(s) after the edit.

*Percentage of total fixations inside the ROI (*percFixInside*).* This percentage is computed after fixating on a ROI after the edit. It is thus independent of *framesToROI*. Different configurations of the ROIs may imply, by nature, different number of fixations inside the ROI. To compensate for this, we compute *percFixInside* relative to the average percentage of fixations inside a ROI, for each ROI configuration, *before* the edit. This metric is indicative of the interest of the viewer in the ROI(s).

*Scanpath error (*scanpathError*).* We compute the RMSE of each scanpath with respect to the corresponding baseline scanpath (see Sec. 6.1). This metric indicates how gaze behavior is altered by the edit; again, we compute this metric after fixating on a ROI after the edit, to make it independent of *framesToROI*.

*Number of fixations (*nFix*).* We compute the ratio between the number of fixations, and the total number of gaze samples after the edit after fixating on a ROI; this way, we eliminate the possible increase in saccades while searching for the ROI after the edit. This metric is therefore indicative of how many fixations and saccades the subject performs. A low value corresponds to a higher quantity of saccades, which in turn suggests a more exploratory behavior, fixating less on any particular region or action.

*State sequences.* We classify users' fixations along time in four different states, corresponding to the ROIs (each clip having one, or two), the background, and a so-called idle state where saccadic eye movements take place and no fixations are recorded. With this classification we are able to describe users' behavior as a state sequence, observing the succession of states with time, as well as the time spent in each of them. In particular, we use a *state distribution analysis* to represent the general pattern of state sequences for each condition, which provides an aggregated view of the frequency of each state for each time interval. We use the R library TraMineR [Gabadinho et al. 2011] for this analysis.

### 6.3 Analysis

Since we cannot assume that our observations are independent, we employ multilevel modeling [Browne and Rasbash 2004; Raudensbush and Bryk 2002] in our analysis, which is well-suited for grouped or related data like ours. Multilevel modeling allows the specification of random effects among the predictors, i.e., it contemplates the possibility that the model might differ for different values of these random effects. In our case, the random effect is the particular subject viewing the stimuli, for which we considered a random intercept.

We include in the regression all four factors ($A$, $E$, $R_b$ and $R_a$), as well as the first-order interactions between them. Since we have categorical variables among our predictors, we recode them to dummy binary variables for the regression. For two of our metrics (*percFixInside* and *nFix*), the effect of the subject was significant ($p = 0.002$ and $p = 0.005$, respectively, in Wald's test), indicating that we cannot treat the samples as independent; we therefore report significance values given by multilevel modeling. For the other two metrics (*framesToROI* and *scanpathError*), the effect of the subject was found to be non-significant ($p = 0.201$ and $p = 0.046$, respectively). Therefore, samples can be considered independent, and we perform factorial ANOVA, together with Bonferroni post hoc analyses to further look for significant effects in our data. Throughout the analysis we use a significance level of 0.01.

*Influence of previous VR experience.* In addition to analyzing the influence of the different factors, detailed below, we also analyze whether the subjects' previous experience using VR had an effect on the results. We record this information in the pre-test questionnaire. None of our subjects used VR frequently, but 69% of them had used VR before at some point. When looking at the effect of this previous VR experience on the metrics employed, we found that it had no effect on any of the metrics tested ($p = 0.600$ for *percFixInside*, $p = 0.832$ for *nFix*, $p = 0.197$ for *framesToROI*, and $p = 0.480$ for *scanpathError*). This is to be expected, since an occasional or rare use of VR is unlikely to cause any change in the results.

*Influence of alignment $A$.* The first thing we observe is that there is a clear effect of the alignment factor on the four dependent variables (metrics) under study. In the case of the *framesToROI* ($F(2, 787) = 198.059, p < 0.001$), the Bonferroni post hoc further shows a significant difference ($p < 0.001$) between all three levels ($A_0$, $A_{40}$ and $A_{80}$). As expected, the further away the ROI is, the longer it takes viewers to find it. Interestingly, the metric suggests an exponential trend with the degrees of misalignment. This is shown in Fig. 6 (left), which includes the goodness of fit, and the 95% confidence interval. Fig. 7 also illustrates this, with strong peaks and larger tails of background fixations after the edit (t = 6 secs.) for $A_{80}$ (bottom row) than $A_0$ (top row).

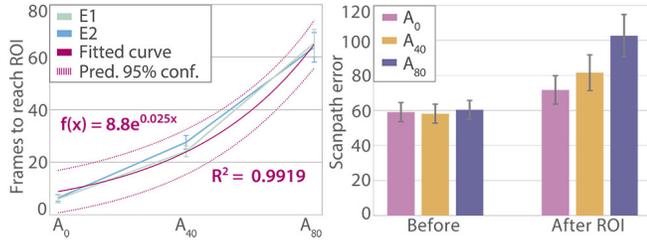

Fig. 6. **Left:** Average *framesToROI* for each alignment. The green and blue curves show average data for the two types of edit ($E_1$ and $E_2$, respectively). We also show a fit to an exponential function, with the associated 95% confidence interval. **Right:** Mean RMSE with respect to the baseline before the edit, and after the edit after seeing the ROI (*scanpathError*) for the different alignment conditions tested. In both plots, error bars show a 95% confidence interval for the mean.

Our *scanpathError* metric ($F(2, 787) = 14.511, p < 0.001$) allows us to dig deeper into this finding, showing in the post hoc analyses that there is no significant difference between $A_0$ and $A_{40}$ ($p = 0.277$), while $A_{80}$ is significantly different to both of them ($p \leq 0.001$ in both cases). This is shown in Fig. 6 (right), comparing directly with the equivalent values before the edit (where, as expected, no significant difference was found). A similar trend can be seen in *percFixInside*: $A_{80}$ is significantly different to $A_0$ ($p < 0.001$), but $A_{40}$ is not ($p = 0.138$).

This effect seems to indicate that the large misalignment alters viewer behavior not only in the time it takes to fixate on the ROI, but also *after* it is found. A closer look reveals that the same significant difference holds for *nFix*: the number of fixations is significantly lower ($p = 0.003$) for $A_{80}$ compared to $A_0$, but this is not the case for $A_{40}$ ($p = 0.954$). This, also shown in Fig. 8 (top) as a radar plot, is a very interesting finding, suggesting that viewers could be more inclined to explore the scene when there is a high misalignment across the edit boundary.

*Influence of type of edit E.* Interestingly, the type of edit (E) has no effect on the fixational behavior after the edit after fixating on the ROI ($p = 0.674$ and $p = 0.430$ for *percFixInside* and *nFix*, respectively). The type of edit did not have a significant effect on *scanpathError* ($F(1, 787) = 0.038, p = 0.846$) either, but the interactions of the type of edit with both ROI configurations did ($p = 0.002$ in both cases). Surprisingly, the type of edit had no significant effect on the *framesToROI* either ($F(1, 787) = 1.373, p = 0.242$), as hinted in Fig. 6.

*Influence of ROI configurations $R_b$ and $R_a$.* We observe no significant influence of these factors on *nFix*, indicating that ROI configuration does not influence the exploratory behavior (how much viewers fixate, in general) of the viewers after the edit once they see one of the ROIs. Interestingly, however, $R_b$ has an effect on *percFixInside*, i.e., on how much viewers fixate on the ROI(s) after the edit after fixating, compared to the total number of fixations in that time period. Note that, while different ROI configurations may imply by nature different number of fixations inside, we are compensating for this effect in the computation of *percFixInside* (Sec. 6.2). Specifically, $R_b$ reveals a difference between two ROIs in the same FOV, and one ROI ($R_{b,1}$ vs. $R_{b,0}$, $p = 0.015$), but not in case of two ROIs in different FOVs ($R_{b,2}$ vs. $R_{b,0}$, $p = 0.792$). This can be seen in Fig. 8 (middle): two ROIs in the same FOV before the edit lead to less fixations on the ROI(s) after the edit. We hypothesize that this is because multiple ROIs before the edit elicit a more exploratory behavior after the edit, in search for more ROI(s) even after having fixated on one.

We also found a significant influence of the ROI configuration after the edit $R_a$ on the deviation of the scanpath wrt. the baseline, *scanpathError* ($F(2, 787) = 168.569, p < 0.001$ for $R_a$); meanwhile, $R_b$ had no significant influence ($F(2, 787) = 1.660, p = 0.191$ for $R_b$). Bonferroni post hocs show that $R_{a,2}$ is significantly different to the other two ($p < 0.001$), while $R_{a,0}$ and $R_{a,1}$ are not significantly different between them ($p = 0.804$). Fig. 8 (bottom) shows this effect: the *scanpathError* is significantly higher for $R_{a,2}$ (two ROIs in different FOVs), indicating that there is more variability in the scanpaths since the two ROIs cannot be looked at simultaneously. Finally, both $R_b$ and $R_a$ had also a significant effect on *framesToROI* ($F(2, 787) = 6.478, p = 0.002$ for $R_b$, and $F(2, 787) = 10.300, p < 0.001$ for $R_a$).

*Other effects.* Additionally, we can observe some new effects in the state distribution sequences (Figs. 7 and 9). In particular, we find an *exploration peak* right at the beginning of each clip, both when the video starts and right after the edit; this peak usually lasts around 1-2 seconds. It is followed by an *attention peak*, again lasting around 1-2 seconds. This effect appears regardless of the ROI configurations and the alignment, and can be observed in Fig. 7. This suggests that users require some time to understand their environment and stabilize their gaze patterns when a change of scenario occurs; after that transitory state, however, their gaze is strongly attracted to the actions being performed (the ROIs).

Last, we analyze more in depth the effect of the two types of edits ($E_1$ and $E_2$) in the particular case of ($R_{b,0}, R_{a,0}$) (edits from one ROI to one ROI). This is one of the simplest cases, but also one of the most relevant, since many current VR film-making strategies are commonly based on a single ROI across scenes. In Fig. 9 we show the *state distribution* for this particular case ($R_{b,0}, R_{a,0}$) for alignments $A_0$ and $A_{80}$, and for the two types of edits. Even though we found no significant effect of the type of edit in our metrics, the graphs suggest a difference that our metrics are not capturing. In particular, it seems that $E_2$ attracts more attention to the ROI after the edit than $E_1$, as seen in the deeper blue valley after the edit in the right column), and this effect is consistent across all alignments. A potential explanation is that the continuity in action before and after the $E_2$ edit acts as an anchor.

## 7 DISCUSSION AND CONCLUSIONS

To our knowledge, our work is the first to attempt a systematic analysis of viewer behavior and perceived continuity in narrative VR content. A systematic exploration of this topic is challenging for two main reasons: (i) the extreme high dimensionality of its parameter space; and (ii) that it involves many discrete, categorical (as opposed to interval or ordinal) variables of influence. Moreover, other basic issues need to be addressed, such as: How does one measure continuity, or viewer behavior? Which are the best metrics to use? Are our observations independent of the subjects? We have relied on the event segmentation theory, which has provided us with some solid ground to carry out our research, and have analyzed previous related studies on traditional cinematography.

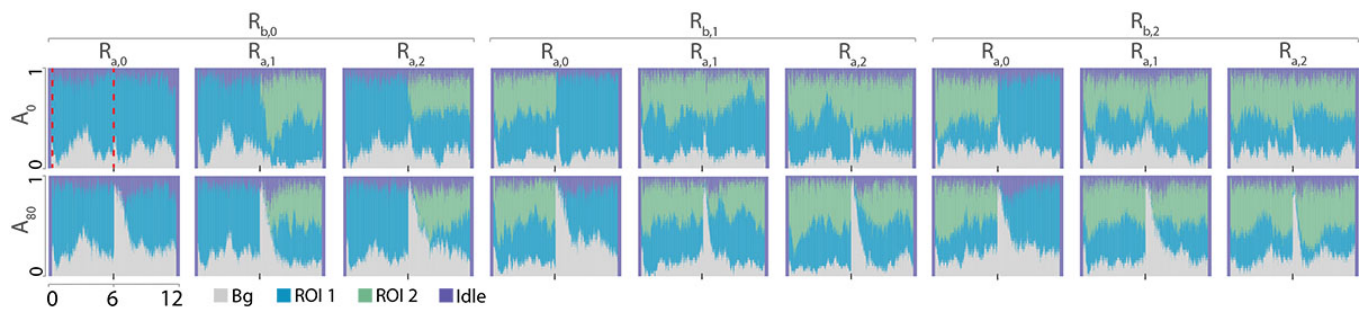

Fig. 7. *State distribution* for all the different combinations of $\mathbf{R_b}$ and $\mathbf{R_a}$, and for alignments $A_0$ (first row) and $A_{80}$ (second row). The different types of edits $E$ are aggregated in each of the aforementioned conditions. The abscissae show time in seconds (the edit takes place at $t = 6$), while the ordinates show aggregated percentage of users. Each plot shows the percentage of users in each state at each time instant. The short Idle periods at the beginning and the end are due to black frames before and after the movie, and thus are not significant to our analysis. The two red lines on the top-left image illustrate the exploration and attention peaks, respectively, reported in the paper.

Our results may have direct implications in VR, informing content creators about the potential responses that certain edit configurations may elicit in the audience. For instance, for a fast-paced action movie our results suggest that ROIs should be aligned across edits, while to evoke a more exploratory behavior, misalignments are recommended. Additionally, from all the narrative 360° movies we have explored, we have found an interesting trend in the number and classification of edits: while in VR movies the great majority of edits are type $E_1$ (action discontinuity), they are by far the least frequent in traditional cinematography, where $E_3$ continuity edits are the most prominent. For example, *The Red Balloon* has 85 continuity edits, 67 spatial/temporal discontinuities, and only 18 action discontinuities. We believe this is due to the immersive nature of narrative VR, where an excessive number of continuity edits would reduce opportunities for free exploration. In the rest of the section, we summarize our main findings, and outline interesting areas of future work ahead.

*Cognition and event segmentation in VR.* We have first replicated an existing cognitive study carried out on the *The Red Balloon* movie, and found many similarities in VR. Like in traditional cinematography, action discontinuities dominate event segmentation in VR, becoming the strongest predictors of event boundaries. Continuity edits do succeed in maintaining the perceived continuity also in VR, despite the visual discontinuity across edit boundaries. This suggests that viewers build a mental model of the shown event structure that is similar to watching a traditional movie, despite the drastically different viewing conditions.

*Measuring continuity effects.* Our analysis has revealed several other interesting findings. Moreover, most of our reported findings have significant values of $p < 0.01$; this minimizes the risk of false positives in our conclusions.

The relation between how misaligned a ROI appears after an edit, and how long it takes viewers to fixate on it, seems to be exponential; this could be used as a rough guideline when performing edits. Even more importantly, large misalignments across edit boundaries do alter the viewers' behavior *even after they have fixated on the new ROI*. A possible interpretation is that the misalignment fosters a more exploratory behavior, and thus could be used to control attention. Two ROIs in the same FOV before an edit seem to elicit a more exploratory behavior as well, even after having located one ROI after the edit.

Other effects not caught by our metrics can be inferred by visual inspection of the state distributions. There seems to be at exploration peak at the beginning of each clip, and a similar attention peak right after the edit, independent of the type of edit. Both suggest that users require some time to adapt to new visual content, before their gaze fixates on ROIs. Also, it appears that the ROI attracts more attention after an $E_2$ edit than after a type $E_1$, perhaps because the consistent action before and after the edit acts as an anchor.

*Limitations and future work.* As in all studies of similar nature, our results are only strictly valid for our chosen stimuli. We have focused on short 360° videos for several reasons: to isolate simple actions, avoiding confounding factors; to gain control over the stimuli, enabling a systematic exploration of the parameter space; and to facilitate the analysis of the gathered data. Some of our findings may therefore not generalize to conditions outside our study.

Of course, many other variables and parameters can be explored in future work, such as other types of cinematographic cuts, longer movies, more complex visual content, the influence of sound, or the effect of fatigue or frequent exposure to VR content. More comprehensive subjective data may also be a valuable source of information, together with our objective gaze data. We believe that the joint study of cognitive mechanisms and cinematographic techniques provides a solid ground to carry out this research.

In summary, we believe that our work is a timely effort, since VR videos are a fast-growing new medium still in its initial exploratory phase, with many content creators testing ways to communicate stories through it. We hope that our findings will be useful as guidelines for VR content creators, especially amateurs, across a reasonable range of situations.

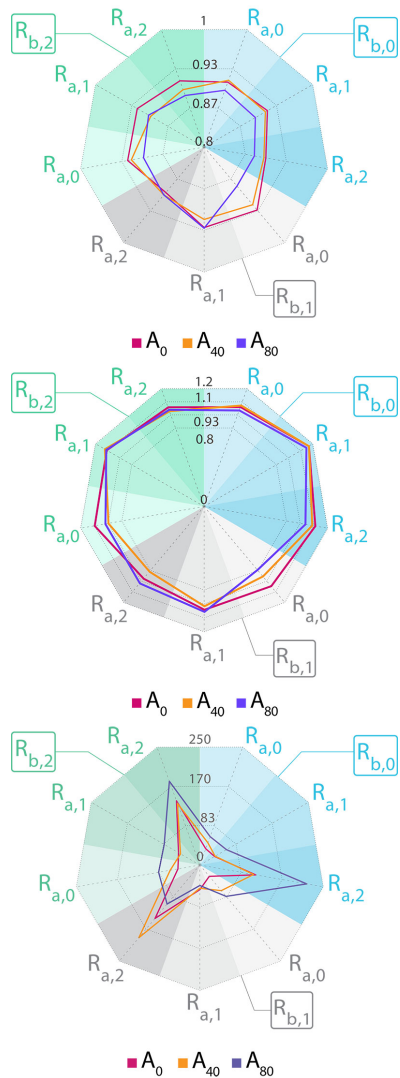

Fig. 8. Radar graphs showing variation of three of our metrics with **A**, **$R_b$** and **$R_a$**. Variation with **E** is not shown. In each graph, the three curves correspond to the three alignment conditions, as labeled in the legend. The radii of the graph correspond to the different combinations between $R_a$ and $R_b$. $R_a$ values are written at each point in the perimeter, while the large colored sectors (blue, gray and green), correspond to $R_b$ ($R_{b,0}$, $R_{b,1}$ and $R_{b,2}$ respectively, as indicated). **Top:** Number of fixations after the edit after fixating on the ROI (*nFix*), which is significantly different for $A_{80}$ than for $A_{40}$ and $A_0$. The scale of the radial axis is enlarged for visualization purposes. **Middle:** Value of *percFixInside* for the different conditions; *percFixInside* is significantly affected by the ROI configuration both before and after the edit (see text for details). **Bottom:** Mean RMSE with respect to the baseline after the edit after fixating on the ROI (*scanpathError*). Please see the text for details.

## ACKNOWLEDGEMENTS

We would like to thank Paz Hernando and Marta Ortin for their help with the experiments and analyses. We would also like to thank *CubeFX* for allowing us to reproduce *Star Wars - Hunting of the Fallen*, and *Abaco digital* for helping us record the videos for creating the stimuli, as well as Sandra Malpica and Victor Arellano for being great actors. This research has been partially funded by an ERC Consolidator Grant (project CHAMELEON), and the Spanish Ministry of Economy and Competitiveness (projects TIN2016-78753-P, TIN2016-79710-P, and TIN2014-61696-EXP). Ana Serrano was supported by an FPI grant from the Spanish Ministry of Economy and Competitiveness. Diego Gutierrez was additionally funded by a Google Faculty Research Award and the BBVA Foundation. Gordon Wetzstein was supported by a Terman Faculty Fellowship, an Okawa Research Grant, and an NSF Faculty Early Career Development (CAREER) Award.

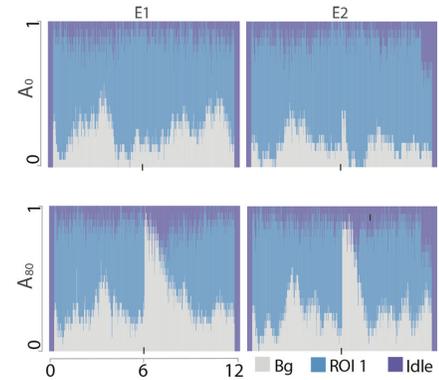

Fig. 9. *State distribution* for ($R_{b,0}$, $R_{a,0}$), for alignments $A_0$ and $A_{80}$, and for the two different types of edits $E_1$ (left) and $E_2$ (right). Although our metrics did not capture this effect, it appears that $E_1$ edits might be harder to understand than $E_2$, as indicated by the deeper blue valley after the edit in the right column.

# APPENDIX

## Stimuli

I. *Definition of cinematographic cut techniques used for the edits*

- Compressed-time cut: It represents the passing of a longer period of time by cutting together key shots (e.g., a character making coffee in a kitchen; while the whole event in real life can last two or three minutes, it can be quickly summarized in a few seconds with a few quick shots showing her pouring water, grinding the coffee, letting it brew, then pouring it into a cup).
- Match-on-action cut: A cut where the second shot matches the action in the first shot (e.g., a character going through a door; as the door starts to open, cut to the character going through the door from the other side).
- Jump cut: Although they need to be avoided in general when shooting the same scene [Arev et al. 2014], they are commonly used to create an abrupt transition from one scene to another.

II. *Movies used for assessing continuity editing in VR*

We used four publicly available movies for carrying out the experiment described in Sec. 4:

- Star Wars - Hunting of the fallen
  https://youtu.be/SeDOoLwQQGo
  Duration: 8:00 minutes.
- Always - A VR story
  https://youtu.be/Tn_V8sVSnoU
  Duration: 5:35 minutes.
- Invisible Episode 2 - Back In The Fold
  https://youtu.be/M3FO3j2z5Tk
  Duration: 4:42 minutes.
- Invisible Episode 5 - Into The Den
  https://youtu.be/M3FO3j2z5Tk
  Duration: 4:05 minutes.

III. *Details on ROI alignment*

We manually aligned the shots of our cuts with Adobe Premiere 2015 CC. In order to do so, we made sure that ROIs were aligned before and after the edits for the $A_0$ condition. Once this edit was generated, we misaligned the second (after-the-cut) shot with respect to the aligned position by 40 degrees for the $A_{40}$ condition, and by 80 degrees for the $A_{80}$ condition. The misalignment was randomly performed to the left or right, but ensuring an equal number of stimuli in each direction. In Fig. 10 we show some examples of different alignments and disposition of the ROIs in our cuts.

## Gaze data processing

We collected gaze points with the eyetracker and head positions with the Oculus DK2. We describe here the main aspects of the processing of this data.

*Gaze scanpaths:* First, we processed the eyetracker samples. We discarded full trials from the eyetracker when the mean confidence for both eyes was lower than 0.6 (the confidence ranges from 0 to 1).

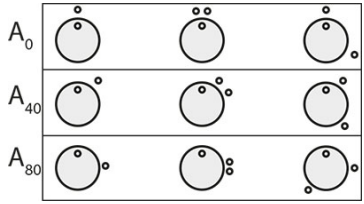

Fig. 10. Examples of different alignments and dispositions of the ROIs in our cuts. **From top to bottom**: Alignments corresponding to 0 degrees for $A_0$, 40 degrees for $A_{40}$, and 80 degrees for $A_{80}$. **From left to right**: Edits aligned from 1 ROI to 1 ROI, to 2 ROIs in the same field of view, and to 2 ROIs in different field of view.

We then linearly interpolated all measurements from the eyetracker whose confidence was below 0.9. Additionally, the head position tracker from the Oculus DK2 has a lower sampling rate than the eyetracker, so in order to match the different sampling rates, we matched each gaze measurement with the closest timestamp of the head position. Second, we matched gaze positions to frames in our videos. Since videos had a frame rate of 60 fps and the eyetracker recorded at 120Hz, 2 gaze positions were recorded per frame. We assigned a single gaze point to each frame by computing the mean of the gaze points corresponding to that frame [Coutrot and Guyader 2014]. Finally, we define a gaze scanpath as the resulting temporal sequence of gaze positions.

*Fixation detection:* We perform fixation detection for our gaze scanpaths with a velocity-based fixation detector. We consider that a gaze point is a fixation when its velocity is below a certain threshold. We calculate this threshold for each scanpath as 20% of the maximum velocity, after discarding the $2^{nd}$ percentile of top velocities [Kübler et al. 2015].

*Outlier rejection:* We discard outliers under two criteria. First, we discard observations when less than 40% of the total number of fixations before the cut occurred inside the ROI. We consider that in such cases users were not paying attention, or did not understand the task. Second, we discard observations that differed significantly from other users' behavior. We do this by following a conservative standard outlier rejection approach, in which an observation is discarded if it fulfils one of the following conditions:

$$\begin{aligned} observation &< (Q_1 - k_d * Q_d) \\ observation &> (Q_3 + k_d * Q_d) \end{aligned} \quad (1)$$

where $Q_1$ and $Q_3$ are the first and the third quartile, respectively; $Q_d = Q_3 - Q_1$; and $k_d = 1.5$.